\documentclass[twocolumn]{IEEEtran} 
\def\BibTeX{{\rm B\kern-.05em{\sc i\kern-.025em b}\kern-.08em
    T\kern-.1667em\lower.7ex\hbox{E}\kern-.125emX}}

\begin{document}
\title{Theoretical Perspectives on Spintronics and Spin-Polarized Transport} 
\author{S. Das Sarma, Jaroslav Fabian, Xuedong Hu, 
and Igor \v{Z}uti\'{c}\thanks
{ This work was supported by the U.S. ONR, DARPA, and the Laboratory
for Physical Sciences.
Department of Physics, 
University of Maryland, College Park, MD 20742-4111,
E-mail: dassarma@physics.umd.edu, jfabian@physics.umd.edu, xdhu@glue.umd.edu,
igor@cooperon.umd.edu}}                                                       

\maketitle

\begin{abstract}
Selected problems of fundamental importance for spintronics and spin-polarized
transport are reviewed, some of them with a special emphasis on their 
applications
in quantum computing and coherent control of quantum dynamics. The role of the
solid-state environment in the decoherence of electron spins is discussed. In
particular, the limiting effect of the spin-orbit interaction on spin
relaxation of conduction electrons is carefully examined in the light of recent
theoretical and experimental progress. 
Most of the proposed spintronic devices involve spin-polarized transport
across interfaces in various hybrid structures.
The specific example discussed here, of a magnetic 
semiconductor/superconductor interface, 
displays many intricacies which a complex spin-dependent interface  
introduces in the spin-polarized transport.
It is proposed that pairs of entangled electrons  in 
a superconductor (Cooper pairs) can be transfered to a non-superconducting
region, and consequently separated for a transport study of the spin
entanglement.
Several important theoretical proposals for quantum computing are based
on electronic and nuclear spin
entanglement in a solid. 
Physical requirements for these proposals to be
useful are discussed and some alternative views are presented. Finally, 
a recent discovery of optical control of nuclear spins in semiconductors
is reviewed and placed in the context of a long-standing search for electronic
control of nuclear dynamics.       
\end{abstract}
 
\begin{keywords}
spintronics, spin-polarized transport, spin coherence,
spin-hot-spot model, spin-based quantum computation,
quantum information, spin entanglement, hybrid semiconductor structures,
spin injection, Andreev reflection.
\end{keywords}
 
\section{Introduction}                                                          

Spintronics 
is a new branch of electronics where electron spin 
(rather than, or, 
in addition to
electron charge) is the active element for information storage and transport
\cite{prinz98,gregg97}. 
Spintronic devices have the potential to
replace and complement various  conventional electronic devices
with improved performance. In a broader sense, spintronics also includes
new fields such as spin-based quantum computation and 
communication \cite{divincenzo95}. 
To determine the feasibility of spintronic devices and more
generally of various applications of spin-polarized transport 
(such as solid-state quantum computing)
it is essential to answer  questions like how to create and 
detect spin-polarized
carriers, how to maintain their spin polarization
and spin coherence, or conversely how 
the spin polarization and spin coherence are destroyed.

In this paper we will explore the question of how 
conduction electrons (and holes) 
lose their spin coherence in view of our recent
theoretical investigation of spin decoherence in 
metals \cite{fabian98,fabian99a,fabian99}.
The
conduction-electron spin relaxation time $T_1$ (which is the same as the
transverse decoherence time $T_2$ for electronic systems with nearly spherical
Fermi surfaces) also determines the quality of spintronic devices.
The longer the conduction electrons remain in a certain
spin state (up or down) the longer and more reliably can they store and
carry information. 

One  important realization  of spintronic devices
is based on hybrid semiconductor structures \cite{das}.
In spite of the initial proposal over three decades ago \cite{aron}
and numerous experimental efforts, one of the key ingredients, the direct
electrical spin injection into non-magnetic semiconductor has only recently
been realized \cite{ham,nat}. Nevertheless by fabricating a
novel class of
ferromagnetic semiconductors \cite{ohno} based on Mn-doped GaAs,
and employing extensive experience
with semiconductor technology which dominates traditional electronics,
a significant progress is expected.
One of the important limitations, however, is the
influence of interfaces between the different materials in such hybrid
structures. In this context, we will discuss here 
spin-polarized transport in hybrid semiconductor structures
including our proposal to study
semiconductor/superconductor structures \cite{zs},
which provide a means to measure the degree of spin polarization and
to investigate the interfacial scattering.

Quantum computation has been one of the most actively studied areas in
general physics in the past few years \cite{divincenzo95}. 
The dream of using quantum
objects such as electrons as the basic unit of a computer, which
is the ultimate of circuit miniaturization, together with the
promise of exponential speed-up due to quantum mechanics, has
drawn intense interest of scientists from a wide range of specialties.
Electron and nuclear spins
are the quantum bits (qubits) of some promising
proposals for quantum computers (QC)
\cite{DL1,BK1,Fumiko,Imam}.  For these proposals
to work, one needs to be able to precisely manipulate the dynamics
of these spins, in particular, to rotate single spins and entangle
two spins.  Here we will focus on  several proposals to produce and to
detect spin entanglement.

From the early days of nuclear magnetic resonance
we know how to manipulate nuclear spins by radio frequency (rf) fields. 
In particular,
rf fields can induce transitions between Zeeman states
and even saturate spin population.
A recent discovery \cite{kikkawa00} claims
to do the same with optical fields, for which nuclear spins are normally
transparent.
Electrons, however, can be spin-polarized by light and transfer
the polarization to nuclei through the hyperfine coupling. By a periodic 
modulation of the
electronic spin population, one can resonantly flip nuclear spins: the
resulting oscillating hyperfine coupling acts as an effective ``rf field.''
Efforts to polarize nuclear spins through the hyperfine coupling are not new.
In the past it was proposed (and in some cases verified) that a nuclear 
polarization can arise from saturating spins of
conduction electrons by a rf field (Overhauser effect \cite{overhauser53b}), or
purely electronically by generating hot carriers (Feher effect \cite{feher59}).
Here we will review some of these proposals and discuss their merits
in the current context of spintronics and coherent control of nuclear
spin dynamics.

\section{Spin decoherence in electronics materials}
 
Conduction electrons lose memory of their spin orientation through
collisions with phonons, other electrons, and impurities. The
crucial interaction which provides the necessary spin-dependent potential
is the spin-orbit interaction. The spin-orbit interaction is a relativistic
effect which can have various sources in electronic materials; the two 
most important sources being the interactions 
between electrons and impurities, and electrons and (lattice host) ions. 
The impurity-induced spin-orbit interaction (Overhauser \cite{overhauser53a})
is a random-site potential and, as such, it can induce momentum scattering
accompanied by spin flip.
The ion-induced spin-orbit interaction is  a different story. This interaction
is nicely periodic and by itself would not lead to any spin relaxation at all.
However, the ion-induced spin-orbit interaction becomes a viable source
of spin relaxation when combined with a momentum scattering mechanism 
(impurities or phonons). This was first realized by Elliott \cite{elliott54}: 
because the periodic lattice potential (that includes a spin-orbit interaction) 
yields Bloch states which are, in general, not spin eigenstates, even a 
spin-independent scattering (by impurities or phonons) can induce spin 
flip \cite{elliott54}. This mechanism of spin relaxation in metals and
semiconductors is now called Elliott-Yafet mechanism (Yafet \cite{yafet63} 
made a significant contribution to the theory by studying
spin-flip electron-phonon interactions). We note that in materials without a 
center of inversion (like GaAs and many other interesting semiconductors), 
there
are other relevant mechanisms of spin-relaxation. These mechanisms along
with recent attempts of modulating spin dynamics in semiconductors are reviewed
in \cite{fabian99}.

Spin relaxation times $T_1$ in metals are typically nanoseconds (the
record, $T_1 \approx 1 \mu$s, is held by a very pure Na sample 
at low temperatures
\cite{kolbe71}). Spin relaxation is an incredibly long process
when compared with momentum relaxation; momentum relaxation times $\tau$ are
just tens of femtoseconds at room temperature. That electron spins are a
promising medium for information storage follows from the large value of
the factor $T_1/\tau$. A crude estimate of $T_1$ is 
$T_1\approx \tau/b^2$, where $b\approx V_{SO}/E_F$, with
$V_{SO}$ denoting an effective strength of the spin-orbit interaction, and
$E_F$ the Fermi energy. Since $V_{SO}\ll E_F$, it follows that
$T_1/\tau\gg 1$. The temperature ($T$)  dependence of $1/T_1$ is similar to 
the temperature dependence
of resistivity $\rho$: At low T (below 20 K) the spin relaxation is 
dominated by impurity scattering and
is temperature independent. At higher temperatures electrons lose
spin coherence by colliding with phonons.  Above the Debye temperature,
where the whole spectrum of phonons is excited and the number of phonons
increases linearly with increasing temperature, $1/T_1\sim T$, similar
to resistivity. In an intrinsic sample (with negligible amount of impurities),
$1/T_1$ follows the Yafet law \cite{yafet63}, $1/T_1 \sim T^5$ (again similar
to $\rho$), which is yet to be seen in experiment.
The case of semiconductors \cite{fabian99} is less clear-cut.
Typical magnitudes of $T_1$ in semiconductors are nanoseconds too, but
$T_1$ varies strongly with magnetic field, temperature, doping, and strain.
The task of sorting different mechanisms at different regimes is
very difficult and remains to be completed \cite{fabian99}.

For how long an electron can travel in a solid-state environment without
flipping
its spin? Is there a limit on $T_1$? In an ideal impurity-free
sample, $T_1$ would approach infinity as temperature gets to the absolute zero.
Thus a recipe to increase $T_1$ at low temperatures is to produce very pure
samples. But the most interesting region is at room temperature. Here phonons
are the limiting factor, not impurities. Since we cannot get rid of phonons,
increasing $T_1$ means reducing the spin-orbit coupling ($b^2$). Typically, 
the heavier is the atom, the stronger is the spin-orbit coupling. 
Therefore lighter metals like Na, Cu, 
or Li have longer $T_1$ than heavy metals like Hg or Pb. 
We do not know how large $T_1$ can be at room temperature, but an educated 
guess would be a microsecond for the materials of current technological
interest.
 
Is there a way to control the spin relaxation rate at least within a few
orders of magnitude? To answer this question we need to understand more
where the strength of the spin-orbit interaction $b$ comes from. We already
pointed out that in general $b\approx V_{SO}/E_F$. This is indeed what a
typical electron on the Fermi surface recognizes as the spin-orbit scattering:
an electron with a spin up in the absence of the spin-orbit interaction 
acquires a spin down amplitude of magnitude $b$ when the interaction is 
turned on. 
Perturbation theory gives $b\approx V_{SO}/\Delta E$, with $\Delta E$ being the 
typical (vertical) energy difference between neighboring bands. For a general 
Fermi surface point $\Delta E\approx E_F$ and one recovers
$b\approx V_{SO}/E_F$. 
But there can be points on the Fermi surface with $\Delta E\ll E_F$! 
Such points occur near Brillouin zone boundaries
or accidental degeneracy lines. In the former case 
$\Delta E\approx V_{G}\ll E_F$,
where $V_G$ is the $G$th Fourier component of the electron-ion interaction
($G$ is the reciprocal lattice vector associated with the Brillouin zone
 boundary); in the latter case $\Delta E$ approaches zero and
degenerate perturbation theory gives $b\approx 1$. We call the points on the 
Fermi surface where $b\gg V_{SO}/E_F$ spin hot spots 
\cite{fabian98,fabian99a,fabian99}. The area of
the Fermi surface covered by spin hot spots is not large, so it may seem that
on the average these points will not contribute much to spin relaxation. 
It turns out \cite{fabian98}, however, that despite
their small weight, spin hot spots dominate the average $b^2$ which is
then significantly enhanced (typically by 1 to 4 orders of magnitude).
Spin relaxation time $T_1$ is correspondingly reduced.

Spin hot spots are ubiquitous in polyvalent metals.
Our theory then predicts that spin relaxation in polyvalent
metals proceeds faster than expected (in fact, the significance
of the points of accidental degeneracy for spin relaxation in Al was first
pointed out by Silsbee and Beuneu \cite{silsbee83}). This is indeed what
is observed. Long before the theory was developed, Monod and Beuneu 
\cite{monod79} collected $T_1(T)$ for different 
metals with the expectation to confirm the formula
$1/T_1(T)\approx b^2/\tau(T)$,
with the simple estimate of $b\approx V_{SO}/E_F$.
This indeed worked for several metals (monovalent alkali and noble metals
like Na or Cu), but not for polyvalent Al, Pd, Be, and Mg (these
remain the only polyvalent metals measured thus far). Spin relaxation times
for the measured polyvalent metals were 2-4 orders of magnitude smaller than 
expected. The explanation of this unexpected behavior came with the 
spin-hot-spot model (see the comparison between the measured and 
calculated $T_1$ 
of Al in \cite{fabian99a}).
 
In addition to providing a theoretical explanation for the longstanding problem
of why electron spins in metals like Al or Mg decay unexpectedly fast, our
theory
also shows a way of tailoring spin dynamics of conduction electrons. Spin hot
spots arise from band structure anomalies which can be shrunk or swollen
by band-structure engineering. Strain, for example, can make a Fermi surface
cross through Brillouin zone boundaries, thus increasing the hot-spot
area and correspondingly $1/T_1$. Other possibilities include alloying,
applying pressure, changing dimensionality of the system, or doping (if dealing
with semiconductors). Any effect that changes the topology of the Fermi surface
will have a severe effect on spin relaxation. This prediction remains to be
verified experimentally. The important result of Kikkawa and 
Awschalom \cite{awschalom99} which shows that 
$T_1$ of some III-V and II-VI semiconductors can be significantly 
(by two orders of magnitude) enhanced by doping, is not 
a manifestation of spin hot spots, but it is still most probably 
(directly or not) a band structure effect.

\section{Spin-polarized transport in hybrid semiconductor structures}

We consider next some  aspects of the spin-polarized transport
in semiconductors and how the studies of  semiconductor/superconductor  
(Sm/S) hybrid structures can be used to investigate the feasibility of novel
spintronic devices. 
With the prospect of making spintronic devices \cite{prinz98,gregg97}
which consist of hybrid 
structures, it is necessary  to understand the influence of interfaces 
between different materials. In the effort to fabricate increasingly smaller
devices, it is feasible to attain a ballistic regime, where the
carrier mean free path exceeds the relevant system size. Consequently,
the scattering from the interfaces plays a dominant
role. In a wide variety of semiconductors the main sources of interfacial 
scattering at the interface with normal metal are arising from the formation 
of a native Schottky barrier \cite{wolf}
and the large difference in carrier densities, 
i.e., Fermi velocity mismatch in the two materials.
In the absence of spin-polarized carriers this leads to reduced
interfacial transparency and different techniques are employed 
to suppress the Schottky barrier which can be examined using the low 
temperature transport measurements in Sm/S structures \cite{belt}.
For a spin-polarized transport in a non-magnetic semiconductor where the 
polarized carriers are electrically injected from a ferromagnet or
ferromagnetic semiconductor, the situation is more complicated.
Magnetically active interface can  introduce both
potential \cite{zs} and the spin-flip scattering leading to
the spin-dependent transmission (spin filtering) across the interface and
the change of the degree of carrier spin polarization.
The latter possibility has profound effect on  spintronic devices
as they rely on the controlled and preferably large carrier spin polarization.

While there are alternative ways available to create spin-polarized carriers
and spin-polarized transport \cite{zs,Fabian} in a semiconductor, an important
obstacle to develop
semiconductor based spintronic devices \cite{das} was to achieve direct
spin injection from a ferromagnet \cite{aron}.
Previous experiments demonstrating spin injection into the  non-magnetic 
metal \cite{john} and into superconductor \cite{gold} have created strong
impetus to advance  studies of spin-polarized transport 
in the corresponding materials.
In the experiment by Hammar {\it et al.} \cite{ham}, 
permalloy (Ni$_{0.8}$Fe$_{0.2}$,Py)
was used as a ferromagnet for the spin injection in two-dimensional
electron gas.
It was theoretically suggested \cite{wee} that limitations for achieving
higher degree of spin polarization are consequences of working in a
diffusive regime and the current conversion near the 
ferromagnet/semiconductor interface. A different approach, which would
circumvent such  difficulties, was proposed by Tang {et al.} \cite{rouk} 
who have considered  spin injection and detection in a ballistic regime.
Subsequent experiments on  spin injection in semiconductors have also
employed diluted magnetic semiconductors 
and  ferromagnetic semiconductors as  sources of spin-polarized
carriers \cite{nat}.
In these cases the effect of reduced interfacial barrier and Fermi
velocity mismatch (as compared to the interface of semiconductor
with metallic ferromagnet) should facilitate injection of
carriers  across the interface
with a substantial degree of spin polarization.
Investigating this point is another reason to perform experiments 
and theoretical studies focusing on the role of interfacial scattering.

We have proposed \cite{zs} employing spin-polarized
transport in Sm/S hybrid structures to address the role of interfacial
scattering and detecting the degree of spin polarization.
Introducing the S region in the semiconductor structures has a dual
purpose. Choosing S as a conventional \cite{tin}, spin-singlet
metallic superconductor (Al, Sn,..)
implies forming of Schottky barrier at the Sm/S interface which 
we want to investigate and by cooling these materials below the temperature
of superconducting transition, $T_c$, scattering processes \cite{and}
present exclusively in the superconducting state can serve as a diagnostic
tool.
At temperatures much lower than $T_c$, and  at low applied bias voltage 
the transport is governed by the
process of Andreev reflection \cite{and}.
Prior to 
work in \cite{zs}, spin-polarized Andreev reflection has been 
investigated theoretically \cite{been} and experimentally \cite{vas2}
only in the context of ferromagnets. 
In this two-particle
process, an incident electron, together with a second electron of the opposite
spin (with their total energy $2E_F$, slightly above and below $E_F$,
respectively) are transfered across the interface into the superconductor
where they form a Cooper pair. Alternatively, this process can be viewed
as an incident electron which at a Sm/S interface is reflected as a
hole belonging to the opposite spin subband, back
to the Sm region while a Cooper pair is transfered to the superconductor.
The probability for Andreev reflection at low bias voltage is thus
related to the square of the normal state transmission coefficient and
can have stronger dependence on the junction transparency than the ordinary
single particle tunneling.
For  spin-polarized carriers, with  different populations in two spin
subbands, only a fraction of the incident electrons from the majority subband
will have a  minority subband partner in order to be  Andreev
reflected.  In the superconducting
state, for an applied voltage smaller than the superconducting gap,
single particle tunneling is not allowed \cite{wolf} 
in the S region and the
modification of the
Andreev reflection amplitude by spin polarization or junction transparency
will be  manifested  in  transport measurements.
To our knowledge, there have not yet been performed experiments
on the spin-polarized transport in Sm/S structures. High sensitivity
to the degree of spin polarization displayed in the experiments
on ferromagnets \cite{vas2}
(including measurements of the spin-polarization for the first time in some 
materials), should serve as a strong incentive
to examine semiconductors in a similar way. 
Performing such experiments in semiconductors
would enable the use of  advanced fabrication techniques, tunable electronic
properties (such as carrier density and the Fermi velocity) and well studied
band structure needed in the theoretical interpretation.

Introducing superconducting regions in the S/Sm structures is not
limited to the diagnostic purpose. They can give rise to new physical 
phenomena relevant to the device operation. For example,
different application of spin-polarized transport in Sm/S structures
has been suggested by Kuli\'{c} and Endres \cite{kul}. They consider properties
of thin films in the ferromagnetic insulator/superconductor/ferromagnetic
insulator (FI/S/FI) configuration which display  qualitatively different 
behavior from the previously studied 
structures where the FI is replaced by 
the metallic ferromagnet (F).
In such F/S/F systems it is known that there are important proximity effects of
extending superconducting order parameter  in the non-superconducting
material. Consequently, it has been shown \cite{rei} that 
they give rise to oscillations in  $T_c$ 
as a function of the thickness of the superconducting region. 
In contrast, for FI/S/FI structures it was shown that
$T_c$ is independent of the thickness of superconducting thin film and  
can be tuned by changing the 
angle  of magnetization direction lying in the planes of each
FI region. It was proposed \cite{kul} that these features
and the simpler physical properties 
compared to the F/S/F systems
can be used to implement switches and logic circuits.  
For example, switching between the normal and superconducting state
could be performed by changing the magnetization directions
(for a spin singlet superconductor $T_c$ depends only
on the relative angle between the two magnetization vectors).
Here we note that
the novel ferromagnetic semiconductors \cite{ohno}
may be suitable candidates for the FI regions discussed above.
With the appropriate Mn doping they would display
insulating behavior and effectively suppress proximity effects. 

\section{Spin-based solid state quantum computation}
For spins in solids to be useful in quantum computing, it
is important that one has certain ways to move information
regarding these spins.  This transfer can be achieved through
nearest (fixed) neighbor interactions, such as among nuclear
spins; or one can use mobile objects like  conduction electrons
in semiconductors.  While the later approach gives us more freedom
in manipulating the system, it is also more susceptible to 
relaxation caused by transport.

One of the first proposal to use electron spins in solids for
the purpose of quantum computing \cite{DL1} suggests confining
electrons in quantum dots.  The spins of trapped electrons serve
as qubits, while quantum dots in which they reside serve as
tags for each qubit.  There is one electron in each quantum dot
so that each qubit can be readily identified.  The individual
electron spins can be easily manipulated by a pulsed local magnetic
field.  It is conceivable that such field can be produced by
local magnetic moments such as a magnetic quantum dot or an STM tip.
Furthermore, if the electrons can be moved 
in the structure
to an area away from the rest of
the qubits, 
without losing its identity, 
the requirement on the magnetic field can be loosened.
Such transport of electrons might be achieved through, for example,
channels \cite{DL2} or STM tips.  Controlled exchange interaction
between electrons in the nearest neighbor quantum dots can
produced desired entanglement between electron spins \cite{DL1,BLD},
while finite magnetic field can be applied to reduce the error rate
during this process \cite{HD}.  It has also been proposed that                  
optical mediated entanglement can be achieved if the quantum dots
are placed in a micro-cavity \cite{Imam}.

To produce a practical electron-spin-based quantum dot QC
is going to be an extremely challenging experimental problem.
For example, because electrons are identical particles,
exchange errors \cite{Ruslai} are always looming whenever
two electrons have wavefunction overlaps \cite{HD}.  Stray
electrons (trapped on surface or impurities) can easily cause
information loss through this channel.  If electronic
qubits are not moved around, single qubit operations would require
precisely controlled local magnetic fields which should not
affect unintended electrons.  Similarly, two-qubit operations
would require well controlled tuning of the gate voltage
between neighboring quantum dots.  Electron spins relax much faster
(in the order of ns to $\mu$s \cite{Awsch}) than nuclear spins
(minutes to hours \cite{BK1}), which would invariably decrease
the signal-to-noise ratio and require error correction.
The above-mentioned problems can be dealt with one by one.
For example, swap gate (or square root of swap) is an essential
ingredient for two-qubit operations.  Thus it would be a big
step forward if one can demonstrate the swap action in a double
dot, even if the swap efficiency is far less than 100\%.  In the
spirit of converting spin information into transport properties,
one  approach might be to inject two streams of electrons
into the two coupled quantum dots, with one stream fully polarized.
By adjusting the speed of injection, one can control the time
that electrons remain in the dots, so that it is possible that                  
at the output the originally unpolarized electron stream would
acquire some degrees of average spin polarization, which can
then be measured.
 
Electron spins are not the only possible building blocks for
proposed spin-based solid state QC.
One proposal that has attracted a lot of attention \cite{BK1}
attempts to combine the extremely long coherence
time for nuclear spins and the immense industrial experience
with silicon processing to produce a scalable QC.
Donor nuclear spins are employed as qubits in this scheme.
Donor electrons also play important roles here.  Controlled
by two types of gates, electrons are used to adjust
nuclear resonance frequency for one-qubit operations and to
transfer information between donor nuclear spins through
electron exchange and hyperfine interaction, 
crucial for two-qubit operations.
The fabrication of regular array of donors may be a daunting task.
The additional ``layer''of the QC structure (the electrons,
as the intermediary) may provide major decoherence channel.
However, despite all  the problems, the exceedingly long life 
time of qubits means that the proposal is one of the more
promising QC models in the long run.

\section{Spin entanglement in solids}

Spin entanglement is an essential ingredient for spin-based
quantum computer, quantum communication, quantum
cryptography, and other applications.
It has been theoretically proposed that two-electron spin
entanglement can be measured using an ordinary electron
beam splitter or through a loop consisted of double-dot
in which electrons undergo cotunneling \cite{SL}.
Another proposal distinguishes singlet and triplet states by
detecting their energy difference \cite{BK2}.   
The common theme here is to measure transport
properties of electrons and infer spin information from
transport.  Direct spin measurement is not impossible with the
current technology (using SQUID), but it is slow and not quite
sensitive enough for the purpose of quantum computing.

Due to the many obstacles we mentioned before in the pursuit
of creating and detecting controlled spin entanglement in solids,
it is useful to separate the two tasks and treat them individually.
For example, to test a detecting scheme, it would be ideal if
we have a well-established source of entangled electrons, so that
we can test the sensitivity and other properties of the detecting
scheme.
Here we propose to use Cooper pairs as such a source.
In many ordinary superconductors, the Cooper pairs are in a singlet
state \cite{tin}. 
Our goal is to transfer a Cooper pair from the superconducting region into
a non-superconducting region as {\it two} spin-entangled electrons
(a Cooper pair injection process 
in analogy to the inverse of previously discussed Andreev reflection).
One conceivable scenario is to use heterostructures with discrete energy 
levels to satisfy energy conservation and to enhance the cross section of 
the process.
If these entangled electrons can be successfully led out of the
superconductor and into a normal region through the above procedure,
they can then be separated
using means such as Stern-Gerlach type of techniques \cite{Fabian}
so that the opposite spin electrons are separated and propagate
in two separate channels.  We thus obtain a source of
two streams of entangled electrons.  By controlling the size of the
point contact between the superconducting and the normal regions,
the arrival-time-correlation between two entangled electrons
can be enhanced so that they can produce signatures of a spin
singlet state.  Indeed, if the Cooper pairs in the superconductor
source are triplets (as suspected for quasi-one-dimensional
organic superconductors
 \cite{q1d}, and for Sr$_2$RuO$_4$ \cite{ruth}), 
signatures of spin triplet state would be present.  To have such
controlled source of entangled electrons would be important
both for testing the entanglement detection schemes and for
applications in areas such as quantum communication.                             

\section{Optical and electronic control of nuclear spin polarization}

The recent discovery by Kikkawa and Awschalom \cite{kikkawa00} of the
optically
induced nuclear spin polarization in GaAs gives an impetus to the search
for new ways of controlling coherent dynamics of nuclear spins.
In the experiment a sample of GaAs held at 5 K 
was placed in a magnetic field of about 5 T. 
Short laser pulses (100 fs) of circularly-polarized light with the frequency
tuned to the GaAs band gap (1.5 eV) were then shot on the sample
(perpendicularly
to the applied field) to create a nonequilibrium population of 
spin-polarized conduction electrons 
(as a result of the circular light polarization). 
The electron
spins then rotated about the total magnetic field which now consisted of the
applied field and whatever field generated by polarized nuclei. By
measuring the rotational frequency, the field induced by the polarized nuclei 
was measured as a function of time. The experiment found that by pumping the 
electron
population with 76 MHz repetition rate laser pulses, the nuclei became 
polarized. After about 250 seconds of pumping the polarization field was 
about 0.1 T. It is not clear what exactly is the mechanism behind the nuclear 
polarization (a simple picture \cite{kikkawa00} based on the Overhauser
effect \cite{overhauser53b} disagrees with the experiment), but there is little
doubt that the polarization is nuclear (because of the large
relaxation times of order minutes) and that it is induced optically.
 
An even more fascinating possibility, also studied by Kikkawa and
Awschalom \cite{kikkawa00}, is a dynamical control of the nuclear
spins. In the standard nuclear magnetic resonance
experiments  nuclear spins which rotate
about an applied field can be flipped by applying a microwave radiation of the
frequency of the spin rotation. This happens because the microwave field has a
component of the oscillating magnetic field perpendicular to the applied
field. But such a perpendicular oscillating field can be created purely
electronically! One just has to create a nonequilibrium population of
spin-polarized electrons with the spin orientation perpendicular to the
applied field, and repeat this process periodically with the period of the 
nuclear
spin rotation. The hyperfine interaction is then the required oscillating
interaction and should be able to resonantly tip nuclear spins. This was
indeed observed \cite{kikkawa00}.  
 
That nuclear spins can be controlled electronically was first suggested by
Feher \cite{feher59} as early as in the late 50's. Feher pointed out that 
nuclear polarization
can be induced by the hyperfine interaction if the effective temperature $T_R$
characterizing the electronic velocity distribution
differs from the electronic spin temperature $T_S$ which determines the 
occupation
of electronic Zeeman states. Feher proposed several mechanisms that would 
lead to
$T_R\ne T_S$ \cite{feher59,clark63}: hot-electron transport, electron drift in 
an electric field gradient or in a perpendicular magnetic field,
and the injection
of electrons whose g-factor differs from the one of the electrons inside
the sample. All these methods rely on the fact that spin equilibration proceeds
slower than momentum equilibration (see Section II). One practical use of this
idea would be a dc-driven maser \cite{clark63}, in which paramagnetic impurities are 
polarized
electronically to an effective negative temperature.                                                                             
We believe that the Feher effect will be revived by new experiments since 
it shows how to manipulate nuclear spins purely electronically (without the
need of either rf or optical fields) which can be of great interest in 
the efforts of integrating standard electronics with quantum information 
processing.


\begin{thebibliography}{1} 
\bibitem{prinz98} G. A. Prinz, Science {\bf 282}, 1660 (1998).
\bibitem{gregg97} J. Gregg {\it et al.}, J. Magn. Magn. Mater. {\bf 175}, 1 
(1997).
\bibitem{divincenzo95} D. P. DiVincenzo, Science {\bf 270}, 255 (1995).
\bibitem{fabian98} J. Fabian and S. Das Sarma, Phys. Rev. Lett.
{\bf 81}, 5624 (1998); J. Appl. Phys. {\bf 85}, 5057 (1999).
\bibitem{fabian99a} J. Fabian and S. Das Sarma, Phys. Rev. Lett. {\bf 83},
1211 (1999).
\bibitem{fabian99} J. Fabian and S. Das Sarma, J. Vac. Sc. Technol. B {\bf 17},
1708 (1999).
\bibitem{das} S. Datta and B. Das, Appl. Phys. Lett. {\bf 56}, 665 (1990). 
\bibitem{aron} A. G. Aronov and G. E. Pikus, Sov. Phys. Semicond. {\bf 10},
698 (1976).
\bibitem{ham} P. R. Hammar, B. R. Bennett, M. Y. Yang, and M. Johnson,
Phys. Rev. Lett. {\bf 83}, 203 (1999).
\bibitem{nat} R. Fiederling {\it et al.}, 
Nature {\bf 402}, 787 (1999);
Y. Ohno {\it et al.}, Nature {\bf 402}, 790 (1999).
\bibitem{ohno} 
J. De Boeck {\it et al.}, Appl. Phys. Lett. {\bf 68}, 2744 (1996);
H. Ohno, Science {\bf 281}, 951 (1998).
\bibitem{zs} I. \v{Z}uti\'c and S. Das Sarma, Phys. Rev. B {\bf 60}, 
R16322 (1999).
\bibitem{DL1} D. Loss and D. P. DiVincenzo, Phys. Rev. A {\bf 57},
120 (1998).
\bibitem{BK1} B. E. Kane, Nature (London) {\bf 393}, 133 (1998).
\bibitem{Fumiko} F. Yamaguchi and Y. Yamamoto, Appl. Phys. A
{\bf 68}, 1 (1999).
\bibitem{Imam} A. Imamoglu, D. D. Awschalom, G. Burkard, D. P.
DiVincenzo, D. Loss, M. Sherwin, and A. Small,
Phys. Rev. Lett. {\bf 83}, 4204 (1999).
\bibitem{kikkawa00} J. M. Kikkawa and D. D. Awschalom, Science {\bf 287},
473 (2000).
\bibitem{overhauser53b} A. W. Overhauser, Phys. Rev. {\bf 92}, 411 (1953).
\bibitem{feher59} G. Feher, Phys. Rev, Lett. {\bf 3}, 135 (1959).
\bibitem{overhauser53a} A. W. Overhauser, Phys. Rev. {\bf 89}, 689 (1953).
\bibitem{elliott54} R. J. Elliott, Phys. Rev. {\bf 96}, 266 (1954).
\bibitem{yafet63} Y. Yafet, in {\it Solid State Physics}, edited by F. Seitz 
and
D. Turnbull (Academic, New York, 1963), Vol. 14.
\bibitem{kolbe71} W. Kolbe, Phys. Rev. B {\bf 3}, 320 (1971).
\bibitem{silsbee83} R. H. Silsbee and F. Beuneu, Phys. Rev. B {\bf 27},
2682 (1983).
\bibitem{monod79} P. Monod and F. Beuneu, Phys. Rev. B {\bf 19},
911 (1979).
\bibitem{awschalom99} D. D. Awshalom and J. M. Kikkawa,
Physics Today {\bf 52}, 33 (1999).
\bibitem{wolf} E. L. Wolf, {\it Principles of electron Tunneling}
(Oxford University Press, Oxford, England 1985).
\bibitem{belt} J. R. Gao {\it et al.},  
Appl. Phys. Lett. {\bf 63}, 334 (1993);
S. De Franceschi {\it et al.},
Appl. Phys. Lett. {\bf 73}, 3890 (1998).
\bibitem{Fabian} J. Fabian and S. Das Sarma, unpublished.
\bibitem{john} M. Johnson and R. H. Silsbee, Phys. Rev. Lett. {\bf 55},
1790 (1985).
\bibitem{gold} A. M. Goldman, V. A. Vas'ko,  P. A. Kraus, and  K. R. Nikolaev,
J. Magn. Magn. Matter. {\bf 200}, 69 (1999).
\bibitem{wee} G. Schmidt, L. W. Molenkamp, A. T. Filip, and B. J. van Wees,
LANL preprint cond-mat/9911014.
\bibitem{rouk} H. X. Tang, F. G. Monzon, R. Lifshitz, M. C. Cross, and 
M. L. Roukes, LANL preprint cond-mat/9907451.
\bibitem{tin} M. Tinkham, {\it Introduction to Superconductivity}
(McGraw-Hill, New York 1996).
\bibitem{and} A. F. Andreev, Sov. Phys. JETP {\bf 19}, 1228 (1964)
[Zh. Eksp. Teor. Fiz. {\bf 46}, 1823 (1964)];
G.E. Blonder, M. Tinkham and T. M. Klapwijk, Phys. Rev. B
{\bf 25}, 4515 (1982).                                                   
\bibitem{been} M. J. M. de Jong and C. W. J. Beenakker, Phys. Rev. Lett.
{\bf 74}, 1657 (1995); J. -X. Zhu, B. Friedman, C. S. Ting,
 Phys. Rev. B {\bf 59}, 9558 (1999);
I. \v{Z}uti\'c and O. T. Valls, Phys. Rev. B {\bf 60}, 6320 (1999);
ibid {\bf 61}, 1555 (2000); S. Kashiwaya, Y. Tanaka, N. Yoshida,
and M. R. Beasley, Phys. Rev. B {\bf 60}, 3572 (1999).
\bibitem{vas2} V.A. Vas'ko, K. R. Nikolaev, V. A. Larkin, P. A. Kraus, and
A. M. Goldman, Appl. Phys. Lett. {\bf 73};
844 (1998); R. J. Soulen Jr. {\it et al.}, Science {\bf 282}, 85 (1998);
S. K. Upadhyay, A. Palanisami, R. N. Louie, and R. A. Buhrman, 
Phys. Rev. Lett.  {\bf 81}, 3247 (1998).
\bibitem{kul} M. L. Kuli\'{c} and M. Endres, LANL preprint cond-mat/9912315.
\bibitem{rei} C. L. Chien and D. H. Reich, 
J. Magn. Magn. Matter. {\bf 200}, 83 (1999).
\bibitem{DL2} D. Loss and D. P. DiVincenzo, J. Magn. Magn. Matter.
{\bf 200}, 202 (1999).
\bibitem{BLD} G. Burkard, D. Loss, and D. P. DiVincenzo, Phys. Rev B
{\bf 59}, 2070 (1999).
\bibitem{HD} X. Hu and S. Das Sarma, LANL preprint quant-ph/9911080.
To appear in Phys. Rev. A.
\bibitem{Ruslai} M. B. Ruskai, LANL preprint quant-ph/9906114.
\bibitem{Awsch}  J. M. Kikkawa and D. D. Awschalom, Nature
{\bf 397}, 139 (1999); Phys. Rev. Lett. {\bf 80}, 4313 (1998);
J. M. Kikkawa, I. P. Smorchkova, N. Samarth, and D. D. Awschalom, Science 
{\bf 277}, 1284 (1997).
\bibitem{SL} G. Burkard, D. Loss, and E. V. Sukhorukov, LANL preprint
cond-mat/9906071; D. Loss and E. V. Sukhorukov, LANL preprint
cond-mat/9907129.
\bibitem{BK2} B. E. Kane {\it et al.}, Phys. Rev. B {\bf 61}, 2961 (2000).
\bibitem{q1d} T. Ishiguro, K. Yamaji, and G. Saito,
{\it Organic Superconductors} (Springer, Berlin, 1998).
\bibitem{ruth} Y. Maeno {\it et al.}, Nature {\bf 372}, 532 (1994).  
\bibitem{clark63} W. G. Clark and G. Feher, Phys. Rev. Lett. {\bf 10}, 134
(1963).
\end {thebibliography}
\end{document}